\documentclass[letterpaper, 10pt, conference]{ieeetran}

\IEEEoverridecommandlockouts  

\usepackage{enumerate}
\usepackage{amssymb}
\usepackage{amsmath}
\usepackage[latin1]{inputenc}
\usepackage{mathrsfs}
\usepackage{psfrag}
\usepackage{graphics}
\usepackage{amsfonts}

\usepackage{epsfig} 
\usepackage{times} 
\usepackage{psfrag}

\input{psfig.tex}

\newtheorem{theorem}{Theorem}

\newtheorem{definition}{Definition}

\newtheorem{remark}{Remark}

\newtheorem{lemma}{Lemma}

\def\salt{\vskip 0.5 true cm}

\input amssym.def
\input amssym

\def\blacksquare{\hbox{\vrule width 4pt height 4pt depth 0pt}}

\def\qedp{\hfill{$\blacksquare$}}

\def\1D#1#2{{{\partial}\over{\partial #2}}#1}

\def\1d#1#2{{{\rm d}\over{{\rm d} #2}}#1}

\def\salt{\vskip 0.2 true cm}

\title{Detecting Topology Variations in Dynamical Networks}

\author{Giorgio Battistelli and Pietro Tesi
\thanks{G. Battistelli is with the
        Dipartimento di Ingegneria dell'Informazione, Universit\'a di Firenze, Via S. Marta 3, 50139 Firenze, Italy
        {\tt\small giorgio.battistelli@unifi.it} \newline \indent  
        P. Tesi is with ENTEG, Faculty of Mathematics and Natural
Sciences, University of Groningen, 9747 AG Groningen, The Netherlands
 {\tt\small p.tesi@rug.nl}}
}

\begin{document}

\maketitle

\begin{abstract}
This paper considers the problem of detecting topology variations
in dynamical networks. We consider a network whose behavior 
can be represented via a linear dynamical system.
The problem of interest is then that of finding 
conditions under which it is possible to detect 
node or link disconnections from prior knowledge of the 
nominal network behavior and on-line measurements. 
The considered approach makes use of analysis 
tools from switching systems theory. A number of results 
are presented along with examples.
\end{abstract}



\section{Introduction}

Recent years have witnessed a growing interest towards
networks of dynamical systems. There is in fact a trend to build modern infrastructures as large-scale networks, which are possibly 
geographically distributed 
\cite{Papachristodoulou}-\nocite{Chakrabortty,Chertkov}\cite{Persis14}. 
Networks of dynamical systems also
arise and play a fundamental role in vehicle formation, cooperative robotics,
surveillance and environment monitoring, to name a few 
\cite{Fiorelli}-\nocite{Beard,Arcak}\cite{Nowzari}.

In many situations, the network behavior is determined or strictly related to its 
underlying topology. This is the case, for instance, in
consensus, coordination and synchronization problems,
where the dynamical system that describes the evolution of the network
is related to the structure of the graph that models 
the interaction among the various networks components \cite{Murray,Scardovi}. 
On the other hand, also the problem of \emph{inferring the network topology from  observations of the network behavior} is of paramount importance, and it is the objective of this paper to explore such a topic. 

Variations of the network topology can have a major impact on stability and/or
performance. For example, in consensus-like networks a 
link disconnection may slow down convergence or even 
destroy agreement when the graph connectivity is lost \cite{Arcak,Murray}.
More importantly, variations of the topology 
may affect the network secure and reliable operation. 
In fact, the strong interdependency among the various elements of the network is such that a failure in one part of the communication infrastructure can rapidly create global cascading effects. This issue is amplified by the fact that failures in the communication infrastructure can be caused not only by equipment failures or human errors but also by intentional attacks \cite{Amin}-\nocite{Sandberg}\cite{Tesi}

This paper considers the problem of inferring variations of the network topology from  observations of the network behavior. Specifically,
we consider a network whose behavior 
can be represented via a linear dynamical system.
The problem of interest is then that of finding 
conditions under which it is possible to detect 
a node or link disconnection from prior knowledge of the 
nominal network topology and measurements of 
the network state or a subset of it.
Contributions to this topic have been recently proposed. 
In \cite{Rahimian1,Rahimian2}, 
the authors address the problem of detecting single and multiple link failures in a multi-agent system under the agreement protocol.
A notion of distinguishable flow graphs is introduced and 
sufficient conditions for achieving distinguishable dynamics
are stated in terms of inter-nodal distances. In \cite{Rahimian3}, the
authors investigate the problem of detection and isolation of link failures by
exploiting the presence of discontinuities in the derivatives of the output responses of a subset of nodes. It is worth noting that the problem of 
inferring variations of the network topology can also be 
addresses by means of topology identification algorithms 
\cite{Franceschelli}-\nocite{Sanandaji}\cite{Kibangou}.
However, identification algorithms do not assume 
prior knowledge of the nominal network topology,
which is possible in many practical circumstances, and,
as such, do not take full advantage of such extra information,
which may be crucial for achieving 
\emph{early} detection of stability and/or performance losses.

The approach taken in this paper makes use of analysis 
tools from switching systems theory. Specifically,  
networks with switching topology can be naturally 
modeled as a switching system, where the switching 
signal determines the current network configuration
(operating mode). Thus, the problem of detecting a
node/link disconnection can be naturally cast as the 
problem of determining under what conditions the 
operating mode of the system can be uniquely reconstructed 
from observations. In the relevant literature, this problem 
is known as the \emph{discernibility}, \emph{distinguishability} 
or \emph{mode-observability} problem 
\cite{ViChSo02}-\nocite{BabEgerHSCC04,BaBaSc01,Lou09,desantis,battSCL}\cite{battAUT}. 

For linear systems, discernibility can be fully characterized 
through simple algebraic conditions. In fact, it is
completely characterized by the eigenspace components related to 
the various operating modes of the system. 
These conditions are generally difficult to refine
because the dynamics related to the various operating modes of the system
need not be related with one another. In the present case,
however, the situation is different because the dynamics 
resulting from a node or link disconnection can be related 
with the nominal one via \emph{interlacing} theorems \cite{Haemers}. 
Moreover, for several graphs of practical relevance, such as 
complete, ring, path and grid graphs, 
an explicit expression for the eigenspace components 
is available.

By exploiting these features, we provide 
necessary and sufficient conditions 
for detecting topology variations for both the cases of 
node and link disconnections. These conditions
are based on simple algebraic tests, which
can be easily checked numerically as well 
as analytically 
whenever an explicit expression for the eigenspace components 
turns out to be available, 
While the analysis is mainly oriented towards a 
theoretical characterization of the detection problem,
the results also provide several insights on how detection 
can be addressed in practice, as well as guidelines 
for the development of sensor placement algorithms.

The remainder of this paper is as follows. In
Section II, we describe the framework of interest and formulate the detection problem. 
In Section III, the main results of the paper are given. Connections with least-square identification are established in Section IV. Finally, Section V ends the paper with concluding remarks.
For convenience, the proofs are reported in the Appendix.

\section{Framework and Problem Formulation}\label{sec:PM}

We consider a network of $n$ nodes, whose  topological 
structure is represented by an undirected graph 
${\mathcal G} := ( {\mathcal V}, {\mathcal E})$,
where ${\mathcal V} := \{1,2,\ldots,n\}$ denotes the node set
and ${\mathcal E} \subseteq {\mathcal V} \times {\mathcal V}$
denotes the edge set. We assume that the network behavior 
can be represented via a linear dynamical system
\begin{eqnarray} \label{eq:nom_system}
\dot x = \Phi x
\end{eqnarray}
where $x \in \mathbb R^{n}$ denotes the network state;
$x_i \in \mathbb R^{n}$, $i \in \mathcal V$, 
denotes the state of the $i$-th network node;  
$\Phi \in \mathbb R^{n \times n}$ is the matrix 
that determines the network behavior. We assume 
that $\Phi = \Phi'$ and that $\phi_{ij} \neq 0$ if and only if
$(i,j) \in \mathcal E$, where $\phi_{ij}$ denotes the 
$(i,j)$-th entry of $\Phi$. 

As a relevant example, consider a classical 
agreement problem in a network of continuous-time integrators with 
local dynamics $\dot x_i = u_i$, which implement a linear \emph{consensus protocol}
with unitary weights,
\begin{eqnarray} \label{eq:consensus}
u_i = \sum_{j \in \mathcal N_i} \left( x_j - x_i \right)
\end{eqnarray}
where $\mathcal N_i$ denotes the set of neighbors of node $i$.
This gives rise to the linear system $\dot x = -L x$,
where $L$ denotes the \emph{graph Laplacian} induced by $\mathcal G$. 
The system is therefore in the same form as 
(\ref{eq:nom_system}) with $\Phi=-L$, \emph{i.e.},
with $\phi_{ij} = 1$ for $j \neq i$ and 
$\phi_{ij} = - | \mathcal N_i |$ for $j = i$. 

\begin{remark}
Although this paper is only concerned with networks whose  topological 
structure is represented by an undirected graph, 
most of the conclusions can be extended to  directed graphs as well. \qedp
\end{remark}

\subsection{Problem formulation}
We regard the pair (${\mathcal G},\Phi)$ as 
representative of the \emph{nominal} behavior of the network. 
The problem of interest is then that of finding 
conditions under which it is possible to detect 
a variation of the network topology via:
i) knowledge of the \emph{nominal} matrix $\Phi$; and
ii) measurements of 
\begin{eqnarray} \label{eq:net_output}
y = M x, \quad M = \textrm{col}\{e_i, i \in \mathcal M \}
\end{eqnarray}
where $e_i \in \mathbb R^n$ denotes the $i$-th versor,
and $\mathcal M \subseteq \mathcal V$ denotes the set of nodes 
whose state is available for measurements.

A variation in the network topology is specified by means of
a pair $(\bar {\mathcal G},\bar \Phi)$, where 
$\bar {\mathcal G}$ describes the novel
topological structure and $\bar \Phi$ describes the
novel network behavior, \emph{i.e.},
\begin{eqnarray} \label{eq:mod_system}
\dot x = \bar \Phi x
\end{eqnarray}
In particular, we assume 
that $\bar {\mathcal G}$ is an undirected 
graph defined as $\bar {\mathcal G} := ( {\mathcal V}, \bar {\mathcal E})$,
where $\bar {\mathcal E} \subset {\mathcal E}$;
$\bar \Phi = \bar \Phi'$ with $ \bar \phi_{ij} \neq 0$ if and only if 
$(i,j) \in \bar {\mathcal E}$, where $\bar \phi_{ij}$ denotes the 
$(i,j)$-th entry of $\bar \Phi$. 
As detailed hereafter, $(\bar {\mathcal G},\bar \Phi)$
captures several scenarios of practical relevance.

\subsubsection{Link disconnection without dynamics reconfiguration}
suppose that the link disconnection affects the nodes $i,j \in \mathcal V$.
Then $\bar {\mathcal G}$ is characterized by  
$\bar {\mathcal E} = \mathcal E  \setminus  \left\{ (i,j),(j,i) \right\}$,
and
\begin{eqnarray}\label{eq:1}
\bar \Phi  = \Phi - \phi_{ij}
\left( e_i e_j'  + e_j e_i' \right)
\end{eqnarray}
In words, the network dynamics remains unchanged 
with the exception that $\bar \phi_{ij} = \bar \phi_{ji}=0$.

\subsubsection{Link disconnection with dynamics reconfiguration} 
this scenario is the same as the previous one with the exception that,
in addition to having $\bar \phi_{ij} = \bar \phi_{ji}=0$, a variation 
occurs also in $\bar \phi_{ii}$ and $\bar \phi_{jj}$. For example,
in the linear consensus problem described above, one has 
$\bar \phi_{ii}=\phi_{ii} + 1$ and $\bar \phi_{jj} + 1$, and
\begin{eqnarray}\label{eq:2}
\bar \Phi  = \Phi + e_i e_i' + e_j e_j' -  e_i e_j'  - e_j e_i'  
\end{eqnarray}

\subsubsection{Node disconnection without dynamics reconfiguration}
Suppose that the node disconnection affects the node $i \in \mathcal V$.
Then $\bar {\mathcal G}$ is characterized by  
$\bar {\mathcal E} = \mathcal E  \setminus  \left\{ (i,j),(j,i); \, j \in \mathcal N_i \right\}$,
while
\begin{eqnarray}\label{eq:3}
\bar \Phi  = \Phi - \sum_{j \in \mathcal N_i } \phi_{ij}
\left( e_i e_j'  + e_j e_i' \right)
\end{eqnarray}

\subsubsection{Link disconnection with dynamics reconfiguration} 
this scenario is the same as the previous one with the exception that
a variation occurs also in $\bar \phi_{ii}$. 
In the linear consensus problem described above, one has 
$\bar \phi_{ii}= 0$, and
\begin{eqnarray}\label{eq:4}
\bar \Phi  = \Phi + \sum_{j \in \mathcal N_i } \left( e_i e_i' + e_j e_j' -  e_i e_j'  - e_j e_i'  \right)
\end{eqnarray}

\section{Main Results}

In this section, we first introduce a notion 
of \emph{discernible networks}. We then 
present the main results of this paper and establish a number 
of connections with several graphs of practical interest. 
To begin with,  notice that the problem of detecting 
a node or link disconnection from measurements 
can be casted as the problem
of finding conditions under which $\Phi$ and $\bar \Phi$
do not give rise to the same dynamics. Networks satisfying 
this property can be therefore referred to as \emph{discernible}. 

We formalize these concepts. \salt

\begin{definition}
A dynamical network is said to described by the pair 
$(\mathcal G, \Phi)$ if $\mathcal G$ is the graph 
describing the network topology and the network 
behavior obeys $\dot x = \Phi x$, where $x$ is the network state.
\qedp
\end{definition}\salt

\begin{definition}\salt
Consider two dynamical networks described by 
$(\mathcal G, \Phi)$ and $(\bar {\mathcal G}, \bar \Phi)$,
respectively. The networks
are said to be \emph{indiscernible} with respect to the state $x_0$
if $e^{\Phi t} x_0 = e^{\bar \Phi t} {x}_0$ for all $t \in \mathbb R_{\geq 0}$
Otherwise, they are said to be discernible. We denote by 
$\mathcal I$ the set of states for which $(\mathcal G, \Phi)$ and $(\bar {\mathcal G}, \bar \Phi)$ 
are indiscernible. \qedp
\end{definition}\salt

\begin{definition}
Given a matrix $M$ as in (\ref{eq:net_output}),
the networks 
are said to be $M$-\emph{indiscernible} with respect to the pair of states $(x_0,{\bar x_0})$
if $M e^{\Phi t} x_0 = M e^{\bar \Phi t} {\bar x}_0$
for all $t \in \mathbb R_{\geq 0}$.
Otherwise, they are said to be $M$-discernible. We denote by 
$\mathcal I(M)$ the set of pairs of states for which $(\mathcal G, \Phi)$ and $(\bar {\mathcal G}, \bar \Phi)$ 
are $M$-indiscernible. \qedp
\end{definition}\salt

Both discernibility and $M$-discernibility can be viewed as particular 
observability problems, which
can be addressed by looking at the parallel interconnection of 
$\dot x =\Phi x$ and $\dot x =\bar \Phi x$. 
For instance, discernibility is equivalent 
to the observability of the pair $(\Delta,\Gamma)$, 
where $\Delta = \textrm{diag}\{\Phi, \bar \Phi\}$ and 
$\Gamma = \left[ I \, -I \right]$,
over the set $\mathcal X = \{ (w,\xi) \in \mathbb R^{2n}: w=\xi \}$.
On the other hand, $M$-discernibility is equivalent 
to the (standard) observability of $(\Delta,\Gamma)$, 
where $\Delta = \textrm{diag}\{\Phi, \bar \Phi\}$ and 
$\Gamma = \left[ I \, -I \right]$.
The latter is the classical condition for  
reconstructing the active mode of a switching linear system from output measurements
\cite{ViChSo02,BabEgerHSCC04}. 
One sees that both discernibility and $M$-discernibility 
depend entirely on 
the eigenspaces of $\Phi$ and $\bar \Phi$, 
which are in general difficult to analyze.  
With respect to the case of switching systems,
however, the analysis here considerably simplifies since $\Phi$ and $\bar \Phi$
comes with a symmetric structure. For clarify, 
we address the cases of discernibility and $M$-discernibility 
separately.

\subsection{Discernibility}

We first consider the discernibility problem. Notice that 
since $\Phi$ and $\bar \Phi$
are symmetric, there exist 
orthonormal matrices $S$ and $\bar S$ such that 
\begin{eqnarray} 
\Phi = S \Lambda S', \quad \bar \Phi = \bar S \bar \Lambda \bar S'
\end{eqnarray} 
with $\Lambda$ and $\bar \Lambda$ diagonal matrices.
Let $\textrm{spec}(\Phi)$ denote the spectrum of $\Phi$. 
Moreover, for any $\lambda \in \textrm{spec}(\Phi)$, let $\mu(\lambda)$
and $V(\lambda)$ denote its multiplicity and eigenspace, respectively.
Finally, let $S(\lambda)$ be the set of columns of $S$ that generate
$V(\lambda)$, \emph{i.e.}, $V(\lambda) = \textrm{span} (S(\lambda))$,
where $\textrm{span}$ denotes the linear span.
We then have
\begin{eqnarray} 
& \displaystyle e^{\Phi t}  = \sum_{\lambda \in \textrm{spec} (\Phi)}
e^{\lambda t}   S(\lambda) S'(\lambda)  \\
& \displaystyle e^{\bar \Phi t}  = \sum_{\lambda \in \textrm{spec} (\bar \Phi)}
e^{\lambda t}   \bar S(\lambda)  \bar S'(\lambda) 
\end{eqnarray} 
From the above expressions, it is straightforward to draw the following conclusions:
\begin{itemize}
\item[i)] If $\textrm{spec} (\Phi) \cap \textrm{spec} (\bar \Phi) = \emptyset$,
then $\mathcal I = \{0\}$, \emph{i.e.} the only indiscernible state 
is the zero state. This is obviously the smallest indiscernibility 
set that one may have.
\item[ii)] Nonzero indiscernible states exist if and only if
there exists some eigenvalue $\lambda$ common to $\Phi$ and $\bar \Phi$
such that $V(\lambda) \cap \bar V(\lambda) \neq \emptyset$, or, equivalently, such that 
${\rm rank} \{ \left [ S(\lambda) \,\, \bar S(\lambda )\right ] \} 
< \mu (\lambda) + \bar \mu (\lambda)$.
\end{itemize} 

Let $\Psi(\lambda)$ be a matrix whose 
columns form an orthonormal basis of $V(\lambda) \cap \bar V(\lambda)$. 
Hence, the set $\mathcal I$ of states for which $(\mathcal G, \Phi)$ 
and $(\bar {\mathcal G}, \bar \Phi)$ 
are indiscernible is given by
\begin{eqnarray} 
\mathcal I = \left\{ x:  x \in \textrm{span} \left( 
\Psi(\lambda), \lambda \in \textrm{spec} (\Phi) \cap \textrm{spec} (\bar \Phi)
\right)  \right\} 
\end{eqnarray}

In view of the above considerations, it is interesting to investigate under which circumstances the variation in the network topology may lead
to a new system matrix $\bar \Phi$ sharing eigenvalue/eigenvector pairs $(\lambda,x)$ with the original system matrix $\Phi$. Clearly, this amounts
to searching for necessary and sufficient conditions for the existence of pairs $(\lambda,x)$  such that
\[
\Phi x = \bar \Phi x = \lambda x 
\]
With respect to the four scenarios of interest described in Section II-A, the following results can be stated, which show that indiscernible states
can be readily inferred by inspection of the components of the eigenvalues of $\Phi$. 

\begin{theorem} (Link disconnection without dynamics reconfiguration).
Consider a disconnection of link $(i,j)$ as in (\ref{eq:1}). 
Then, the networks are indiscernible with respect to a state 
$x \in V(\lambda)$, with $\lambda \in {\rm spec} (\Phi)$, if and only if $x_i = x_j = 0$. \qedp
\end{theorem}

\begin{theorem} (Link disconnection with dynamics reconfiguration).
Consider a disconnection of link $(i,j)$ as in (\ref{eq:2}). 
Then, the networks are indiscernible with respect to a state 
$x \in V(\lambda)$, with $\lambda \in {\rm spec} (\Phi)$, if and only if $x_i = x_j$. \qedp
\end{theorem}

\begin{theorem} (Node disconnection without dynamics reconfiguration).
Consider a disconnection of node $i$ as in (\ref{eq:3}). 
Then, the networks are indiscernible with respect to a state 
$x \in V(\lambda)$, with $\lambda \in {\rm spec} (\Phi)$, if and only if
\begin{equation}\label{eq:prop3}
x_i \sum_{j \in \mathcal N_i} \phi_{ij}  = 0 , \quad \sum_{j \in \mathcal N_i} \phi_{ij} x_j = 0 .
\end{equation}
If, in addition, we assume that $\phi_{ij} \ge 0$ for any $i \ne j$, 
then condition (\ref{eq:prop3}) becomes $x_i = 0$ and $x_j = 0$ for any 
$j \in \mathcal N_i$. \\ \mbox{ }\qedp
\end{theorem}

\begin{theorem} (Node disconnection with dynamics reconfiguration).
Consider a disconnection of node $i$ as in (\ref{eq:4}). 
Then, the networks are indiscernible with respect to a state 
$x \in V(\lambda)$, with $\lambda \in {\rm spec} (\Phi)$, if and only if
\begin{equation}\label{eq:prop4}
x_j = x_i , \quad \forall j \in \mathcal N_i \, .
\end{equation}
If, in addition, we assume that $\Phi = -L$ (as in the linear consensus protocol (\ref{eq:consensus})) and we consider a non-null Laplacian eigenvalue $\lambda$,
then condition (\ref{eq:prop3}) becomes $x_i = 0$ and $x_j = 0$ for any $j \in \mathcal N_i$.
\qedp
\end{theorem}

From the above results, it can be seen  that, in general, 
node disconnections are easier to detect than link disconnections.
Similar considerations can be made to conclude that
a link/node disconnection without dynamics reconfiguration is easier to detect as compared to a disconnection
with dynamic configuration. 

As an example, consider again the case
$\Phi = -L$, where $L$ is the graph Laplacian of $\mathcal G$. 
As well known, the all-ones vector 
$\mathbf{1}$ is always an eigenvector of $L$ associated to the eigenvalue $0$.
After a link/node disconnection with dynamic reconfiguration, the novel dynamic matrix $\bar \Phi $ will coincide with $-\bar L$, where $\bar L$ is the Laplacian of the graph $\bar { \mathcal G }$.
Hence, the all ones vector $\mathbf{1}$ will be an eigenvector associated to the $0$ eigenvalue also for $\bar \Phi = -\bar L$. Hence, in this case, the stationary state $x = \mathbf{1}$
turns out to be indiscernible for any link/node disconnection. 
This is consistent with the fact that $x = \mathbf{1}$ 
satisfies the conditions of Theorem 2 and 4 for any $i$ and any $j$,
being all its components identical. 
On the contrary, when a link/node disconnection without dynamic reconfiguration occurs,  $x = \mathbf{1}$ is no longer an eigenvector of $\bar \Phi$ 
and, hence, is discernible 
(indeed $x = \mathbf{1}$ does not satisfy the conditions of Theorem 1 and 3).
If we further assume that the graph $\mathcal G$ is connected, then $\mathbf{1}$ turns out to be the unique eigenvector with eigenvalue $0$. Then, the second part of Theorem 4 can be
exploited to conclude that, with the exception of $x = \mathbf{1}$, the two cases of  node disconnection with or without dynamic reconfiguration give rise to the same 
indiscernible eigenvectors.

\subsection{$M$-Discernibility}

We now turn the attention to the problem of detecting 
a topology variation from observations of an output vector 
$y = M x$. In this respect, notice preliminarily
that a state $x \in \mathcal I$ for which $(\mathcal G, \Phi)$ and $(\bar {\mathcal G}, \bar \Phi)$ are indiscernible does always generate indiscernible output trajectories.
Then, if we define the set $\mathcal I_{\rm P} = \{(x,x): \, x \in \mathcal I\}$, 
we have that $\mathcal I_{\rm P} \subseteq \mathcal I (M)$ 
irrespective of the choice of the output matrix $M$. 
On the other hand, by letting $M=I$ (\emph{i.e.}, by observing all the network nodes)
we clearly have $\mathcal I_{\rm P}  = \mathcal I (I)$,
where $I$ stands for the identity matrix of an appropriate dimension.
Hence, a first important problem is how to choose the matrix $M$ so that 
$\mathcal I_{\rm P}  = \mathcal I (M)$.
This amounts to asking where sensors nodes should be placed in order to 
guarantee that discernibility implies $M$-discernibility. 
When such a condition holds, we say that the matrix $M$ ensures {\em output discernibility}.
Since we have
\begin{eqnarray*} 
& M \displaystyle e^{\Phi t}  = \sum_{\lambda \in \textrm{spec} (\Phi)}
e^{\lambda t}   M S(\lambda) S'(\lambda)  \\
& M  \displaystyle e^{\bar \Phi t}  = \sum_{\lambda \in \textrm{spec} (\bar \Phi)}
e^{\lambda t}   M \bar S(\lambda)  \bar S'(\lambda)  
\end{eqnarray*} 
the following result follows at once. 

\begin{theorem}
Consider an observation vector $y$ as in (\ref{eq:net_output}). Then, 
condition $\mathcal I_{\rm P}  = \mathcal I (M)$ holds
if and only if the following conditions are satisfied:
\begin{enumerate}[(i)]
\item ${\rm rank} \left \{ M S(\lambda) \right \} = \mu (\lambda)$ for any $\lambda \in {\rm spec} (\Phi) \setminus {\rm spec} (\bar \Phi)$;
\item ${\rm rank} \left \{ M \bar S(\lambda) \right \} = \bar \mu (\lambda)$ for any $\lambda \in {\rm spec} (\bar \Phi) \setminus {\rm spec} (\Phi)$;
\item ${\rm rank} \left \{ M \left [  S(\lambda) \; \bar S(\lambda) \right ] \right \} = 
{\rm rank} \left \{ \left [  S(\lambda) \; \bar S(\lambda) \right ] \right \}$ for any 
$\lambda \in {\rm spec} (\Phi) \cap {\rm spec} (\bar \Phi)$.
\end{enumerate}
\qedp
\end{theorem}

Notice that condition (i) amounts to requiring that all the states belonging to $V (\lambda)$, 
with $\lambda$ eigenvalue of $\Phi$ but not of $\bar \Phi$,
are observable from the output $y = M x$. The same property 
is required by condition (ii) for all the eigenvalues of $\bar \Phi$, which are not eigenvalues of $\Phi$.
Finally, condition (iii) amounts to requiring that, for any eigenvalue $\lambda$ that is shared by $\Phi$ and $\bar \Phi$, and for any $(x , \bar x) \in V (\lambda) \times \bar V(\lambda)$,
one has $M x = M \bar x$ if and only if $x = \bar x$.

Building upon Theorem 5, an  
expression for $\mathcal I(M)$ can be given. To this end, for any $\lambda \in {\rm spec} (\Phi) \setminus {\rm spec} (\bar \Phi)$, let 
$K (\lambda , M)$ be a matrix whose columns form a basis of the linear space $\{ (x,0) \in \mathbb R^{2n}: \, x \in V(\lambda) \mbox{ and }  M x = 0\}$. Let $\bar K (\lambda , M)$ be defined in a similar
way with respect to $\lambda \in {\rm spec} (\bar \Phi) \setminus {\rm spec} (\Phi)$. Finally, for any $\lambda \in {\rm spec} (\Phi) \cap {\rm spec} (\bar \Phi)$, 
let $\Upsilon (\lambda, M)$ be a matrix whose columns form a basis of the linear space $\{ (x,\bar x) \in \mathbb R^{2 n}: \, x \in V(\lambda), \bar x \in V (\lambda), \mbox{ and } 
M x = M \bar x \}$. Then, we have
{\setlength\arraycolsep{0pt} 
\begin{eqnarray*} 
\mathcal I (M) \, &=& \,
\textrm{span} \left \{ 
 \Upsilon (\lambda,M), \lambda \in \textrm{spec} (\Phi) \cap \textrm{spec} (\bar \Phi) \right \} 
    \\
&&\cup \, \textrm{span} \left \{ 
K(\lambda,M), \lambda \in \lambda \in {\rm spec} (\Phi) \setminus {\rm spec} (\bar \Phi)
\right \} \\
&&\cup \,  \textrm{span} \left \{ 
\bar K(\lambda,M), \lambda \in \lambda \in {\rm spec} (\bar \Phi) \setminus {\rm spec} (\Phi) \right \}
\end{eqnarray*}}%

Theorem 5 provides interesting insights on the 
\emph{number of sensors} needed so as to have output discernibility. 
Consider, for example, the ideal situation
in which $\Phi$ and $\bar \Phi$ are discernible from all the states, 
\emph{i.e.}, $\mathcal I = \{ 0 \}$. Then, condition (iii) becomes 
\[
{\rm rank} \left \{ M \left [  S(\lambda) \; \bar S(\lambda) \right ] \right \} =  \mu (\lambda) + \bar \mu (\lambda)
\]
for any $\lambda \in {\rm spec} (\Phi) \cap {\rm spec} (\bar \Phi)$. Notice also that the rank in the left-hand side cannot exceed ${\rm rank} \{ M \}$, which,
in turn, is equal to the number of measured nodes. Then, we can conclude that, in order to have output
discernibility, one needs a number of sensors at least equal to the maximum among $\mu (\lambda)$ for  $\lambda \in {\rm spec} (\Phi) \setminus {\rm spec} (\bar \Phi)$,
$\bar \mu (\lambda)$ for $\lambda \in {\rm spec} (\bar \Phi) \setminus {\rm spec} (\Phi)$, and  $\mu (\lambda) + \bar \mu (\lambda)$ for $\lambda \in {\rm spec} (\Phi) \cap {\rm spec} (\bar \Phi)$.
Note that this is just a lower bound, since Theorem 5 does not exclude that a larger number of sensors may be needed. Nevertheless, such considerations indicate that,
similar to what happens when standard observability is addressed 
\cite{Notarstefano}, the number of nodes that should be available 
for measurements increases with the multiplicity
of the eigenvalues.

As for the \emph{sensor placement}, one can see that in order to satisfy condition (iii) the sensors $i \in \mathcal M$ must be positioned so that the rows of the matrix $\left [  S(\lambda) \; \bar S(\lambda) \right ] $
corresponding to the indices $i \in \mathcal M$ contain at least one non-zero minor of order ${\rm rank} \left \{ \left [  S(\lambda) \; \bar S(\lambda) \right ] \right \} $.
Analogous considerations can be given for conditions (i) and (ii).

In particular, condition (ii) becomes tricky when, starting from a connected graph $\mathcal G$,  a topology variation gives rise to multiple connected components in the graph $\bar{ \mathcal G}$
(notice that this always happens in the case of node disconnection). 
Specifically, let $\bar{ \mathcal G}$ consist of $N$ mutually disjoints components
$\bar{ \mathcal G}^1, \ldots, \bar{ \mathcal G}^N$, and let $\mathcal N^{k}$ be the set of nodes belonging to $\bar{ \mathcal G}^k$ 
(clearly $\sum_{k=1}^N  |\mathcal N^{k}|  = n$). Then, as well-known, 
\[
{\rm spec} (\bar{ \mathcal G}) = \bigcup_{k=1}^N {\rm spec} (\bar{ \mathcal G}^k)
\]
and, in addition, for any $\lambda \in {\rm spec} (\bar{ \mathcal G}^k) $ there exist eigenvectors $x \in \bar V(\lambda)$ such that $x_i \ne 0$ if and only if $i \in \mathcal N^k$. As a consequence,
it is immediate to verify that condition (ii) can be satisfied only by placing sensors in each one of the mutually disjoints components
$\bar{ \mathcal G}^1, \ldots, \bar{ \mathcal G}^N$.

Clearly, this latter requirement can be quite restrictive in practice. For instance, this implies that one can have output discernibility with respect to any possible node disconnection only
by placing a sensor in each network node. Hence, instead of requiring complete output discernibility, in many situations it may be of interest to restrict the attention only to some of the connected components
of the graph $\bar{ \mathcal G}$. This can be done in a straightforward way by considering in condition (ii) only the eigenvalues and eigenvectors pertaining to the connected components of interest.
For example, in the case of disconnection of node $i$, one can restrict the attention to the component with node set ${\mathcal N} \setminus \{ i \}$ by excluding from condition (ii) the eigenvector $e_i$ pertaining to the
trivial component $\{ i \}$.

\section{A least-squares criterion for detection of topology variations}

In the previous section, we have provided conditions
under which it is theoretically possible to detect 
a topology variation by observing the evolution of  the state $x_i$ in a subset $\mathcal M$ of the
network nodes $\mathcal N$. From a practical point of view, this can be done by resorting to a least-squares criterion, as detailed hereafter.

Suppose that, starting from time $t_0$, a certain number, say $N$, of samples of the output vector $y$ are collected at the time instants
$t = t_0 + k T$ for $k = 0, \ldots, N-1$, where $T \in \mathbb R_{> 0}$.  In particular, to account for the possible presence
of a measurement noise, let each sample $z_k$ be of the form 
\begin{eqnarray} \label{eq:data}
z_k = y (t_0 + k T) + v_k
\end{eqnarray}
with $v_k$ an unknown but bounded discrete-time noise signal. We assume that an upper bound 
$E_v$ on the energy of the sequence $\{ v_k\}$ is known, 
\emph{i.e.}, 
$(\sum_{k=0}^{N-1} \| v_k \|^2)^{1/2} \le E_v$ where
$\| \cdot \|$ stands for Euclidean norm. 
Hereafter, the vector of all the collected samples will be
denoted by
\[
Z_N = {\rm col} (z_0, \ldots, z_{N-1}) 
\] 
\begin{remark}
It is worth noting that (\ref{eq:data}) amounts to making use of synchronous measurements.
While this hypothesis may be restrictive in some cases, 
there are many applications where the measurement devices are equipped with global positioning system (GPS) units. This is the case, for instance, in many smart grid applications
where Phasor Measurement Units are sampled from widely dispersed locations 
and synchronized via a common GPS reference  \cite{Chakrabortty}. \qedp
\end{remark}

Let $\mathcal O_N$ and $\bar{\mathcal O}_N$
denote the sampled-data observability matrices associated with $\mathcal G$ and $\bar{\mathcal G}$, respectively.
Clearly, we have
\begin{equation} \label{eq:obs_matrix}
\mathcal O_N  := 
\left(
\begin{array}{c}
\, M \,\\
\,M\, e^{\Phi \, T} \, \\
\vdots \\
\,M\, e^{\Phi \, (N-1) \, T}  
\end{array}
\right ) , \,\, \bar {\mathcal O}_N  := 
\left(
\begin{array}{c}
\, M \,\\
\,M\, e^{\bar \Phi \, T} \, \\
\vdots \\
\,M\, e^{\bar \Phi \, (N-1) \, T}  
\end{array}
\right ) 
\end{equation}
Notice now that, when the state evolution  is generated by the nominal network $(\mathcal G, \Phi)$, the sampled outputs are of the form $Z_N = \mathcal O_N x(t_0) + V_N$ where $V_N = {\rm col} (v_0, \ldots, v_{N-1})$ and $x(t_0)$ is the (unknown)
state at time $t_0$. 
Then, the least-squares cost function
\[
\pi (Z_N) = \min_{x \in \mathbb R^n} \|Z_N - \mathcal O_N \, x \| 
\]
provides a quantitative measure of how close the observed output behavior is to the nominal ones. In fact,
whenever the output samples $Z_N$ arises from the nominal network, we have $\pi (Z_N) \le E_v$. 

Similarly, any output behavior generated by $(\bar{ \mathcal G}, \bar \Phi)$ leads to sampled outputs of the form $Z_N = \bar {\mathcal O}_N x(t_0) + V_N$ and, hence, the least-squares cost function
\[
\bar \pi (Z_N) = \min_{x \in \mathbb R^n} \|Z_N - \bar{ \mathcal O}_N \, x \| 
\]
provides a quantitative measure of the distance between the observed outputs and 
the set of behaviors associated with the modified topology. 

Then, by computing the quantities  $\pi (Z_N)$ and $\bar \pi (Z_N)$, the following conclusions can be readily drawn:
\begin{enumerate}[(a)]
\item  when $\pi (Z_N) > E_v$, the output samples are not consistent with the nominal network; hence we can conclude
that a variation from the nominal behavior has occurred.
\item when $\pi (Z_N) \le E_v$ and $\bar \pi (Z_N) > E_v$, the output samples are consistent only with nominal behavior; hence
we can exclude the variation associated with $(\mathcal G, \Phi)$. 
\item when both $\pi (Z_N) \le E_v$ and $\bar \pi (Z_N) \le E_v$, we cannot conclude since the sampled outputs are consistent
with both the nominal and the modified behavior.
\end{enumerate}
Of course, the considered framework can be easily 
extended so as to account for different possible topological variations.
In particular, the detection of a topological variation 
as in case (a) requires only the computation of the cost function
associated with the nominal behavior. On the other hand, 
a ``validation" test as in case (b) requires, in
general, the computation of one  cost function for each 
possible variation. Further, also in case (a), computation of
the other cost functions can be useful in order to possibly 
identify the topological variation.

As for case (c), this corresponds to the situation 
in which the information contained in sampled outputs is not
sufficient to conclude on the underlying topology. 
Clearly, this case may arise when the measurement noise 
is sufficiently large so as to mask the data information.
However, case (c) is also inherently connected to the concept of 
$M$-indiscernible states as previously defined. In fact, let $\underline {\mathcal I} (M)$ and $\bar {\mathcal I} (M)$ be the projections on the first and, respectively, last $n$ components of the set $\mathcal I (M)$, \emph{i.e.},
\[
\begin{split}
\underline {\mathcal I} (M) = \{ x \in \mathbb R^n : \exists \bar x \in \mathbb R^n \mbox{ such that } (x,\bar x) \in \mathcal I (M) \} \\
\bar {\mathcal I} (M) = \{ \bar x \in \mathbb R^n : \exists  x \in \mathbb R^n \mbox{ such that } (x,\bar x) \in \mathcal I (M) \} 
\end{split}
\]
Then, it is an easy matter to see that when the output 
behavior is generated by the modified network
starting from state $x(t_0) \in \bar {\mathcal I} (M) $, 
one has $\pi (Z_N) \le E_v$ and $\pi (Z_N) \le E_v$
since both $\pi (\bar {\mathcal O}_N x(t_0) ) = 0$ and $\bar \pi (\bar {\mathcal O}_N x(t_0) ) = 0$.
In this respect, it is worth noting that
although $\bar {\mathcal I} (M)$ is defined with respect to an ideal situation  
(\emph{i.e.}, assuming to measure the noise-free continuous-time evolution of 
$y = M x$), it turns out that such a set
plays a fundamental role also in the practical situation of 
sampled outputs affected by measurement noises, provided
that the sampling is non-pathological \cite{kreisselmeier}.

\begin{lemma} Let the state trajectory be generated by the modified network $(\bar{\mathcal G}, {\bar \Phi})$. Furthermore, suppose that
for any $\lambda, \bar \lambda \in {\rm spec} (\Phi)  \cup {\rm spec} (\bar \Phi)  $ with $\lambda \ne \bar \lambda $
the following condition holds
{\setlength\arraycolsep{0pt} 
\begin{eqnarray}\label{eq:obs:sample}
{\rm Im}(\lambda-\bar \lambda) 
\, &\ne& \, \frac{2 \pi h}{T } \mbox{ for } h \in \mathbb Z \setminus \{0\} 
\nonumber \\
&& \quad  \mbox{ whenever } {\rm Re} (\lambda-\bar \lambda) = 0 \, .
\end{eqnarray}}%
Finally, let $N \ge 2 n$.
Then, case (c) can occur only if
\begin{equation}\label{eq:dist}
d (x(t_0),  \bar {\mathcal I} (M)) \le \alpha E_v
\end{equation}
where $\alpha$ is a suitable positive constant and $d (\cdot, \cdot)$ stands for 
point-set distance. \qedp
\end{lemma} 

In view of Lemma 1, it can be seen that, when the state $x (t_0)$ is far enough from the set of $M$-indiscernible states 
(in the sense that condition (\ref{eq:dist}) does not hold)
and the sampling is non-pathological, then the sampled outputs provide sufficient information for detecting
that a topological variation has occurred.  Bounds on the constant $\alpha$ can be found, 
as in \cite{battSCL},
in terms of the cosine of the smallest non-null angle between the linear subspaces 
${\rm span} (\mathcal O_N)$ and ${\rm span} (\bar{\mathcal O}_N)$. Notice finally that condition
(\ref{eq:obs:sample}) is nothing but the well-known Kalman-Bertram criterion for the
observability of sampled-data systems applied to the pair $(\Gamma, \Delta)$.

Notice that a result analogous to Lemma 1 can be derived also 
in the case of output behaviors generated by the nominal network 
$({\mathcal G}, \Phi)$, with the set $\bar {\mathcal I} (M)$ replaced by $\underline {\mathcal I} (M)$. 

\section{An Example}

In order to illustrate some of the previous results in an easy 
manner, we consider the simple, yet non trivial, case of a linear consensus protocol with unitary weights over the $n$-dimensional path graph $\mathcal G = P_n$
with dynamics reconfiguration, which is standard 
in consensus-like algorithms. We focus on the 
discernibility problem.
Recall that the Laplacian $L$ of $P_n$ has eigenvalues 
$\lambda^{[k]} = 2 - 2cos(\pi k /n)$ and eigenvectors
\begin{eqnarray*}
&& x^{[k]}_i = cos(\pi k i /n - \pi k/2n) \\ 
&& \quad \quad \quad k \in \{0,1,\ldots,n-1\}, i \in \{1,2,\ldots,n\}
\end{eqnarray*}
where 
$\lambda^{[k]}$ denotes the $k$-th eigenvalue and
$x^{[k]}_i$ denotes the $i$-th component of the $k$-th eigenvector.
This describes the nominal network $(\mathcal G,-L)$.

\begin{figure}[tb]
\begin{center}
\psfrag{0}{{\small $\substack{\\ \\ 0}$}}
\psfrag{pim}{{\small $\substack{\\ \\ \frac{\pi}{2}}$}}
\psfrag{pi}{{\small $\substack{\\ \\  \pi}$}}
\psfrag{piM}{{\small $\substack{\\ \\ \frac{3\pi}{2}}$}}
\psfrag{2pi}{{\small $\substack{\\ \\ 2\pi}$}}
\includegraphics[width=0.47 \textwidth]{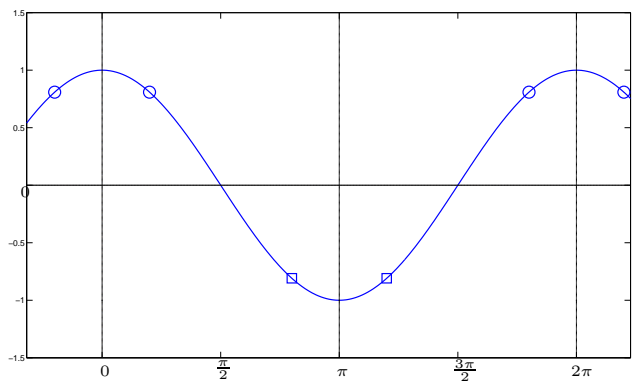} 
\linespread{1}\caption{Indiscernible points for the path graph in case 
of link disconnections. Letting $A := \pi k i /n - \pi k/2n$ and  
$B := \pi k i /n + \pi k/2n$, one has $cos(A)=cos(B)$ whenever 
$A= \pm m \pi -\Delta/2$ and $B=A+\Delta$, where 
$m \in \mathbb N$ and $\Delta \in [0,\pi)$.
} 
\label{fig:cos}
\end{center}
\end{figure}

We consider first the case of disconnection of a link $(i,j)$.
Notice that we can restrict the attention to the situation where $j=i+1$
since the case $j=i-1$ is specular. In accordance with Theorem 2, the
link disconnection is not detectable if and only if
{\setlength\arraycolsep{0pt} 
\begin{eqnarray*} 
cos(\pi k i /n - \pi k/2n) \, &=& \,  cos(\pi k j /n - \pi k/2n) \\
&=& \,  cos(\pi k i /n + \pi k/2n)
\end{eqnarray*}}%
for some $k \in \{0,1,\ldots,n-1\}$ and $i \in \{1,2,\ldots,n-1\}$.
For convenience, let us define 
$A := \pi k i /n - \pi k/2n$ and  $B := \pi k i /n + \pi k/2n$.
By looking at the cosine function (\emph{cf.} Figure \ref{fig:cos}),
one sees that discernibility is violated when $A$ and $B$
take the form $A = \pm m \pi -\Delta/2$ and $B = A +\Delta$,
where $m \in \mathbb N := \{0,1,2,\ldots\}$ and $\Delta \in [0,\pi)$.
The reason for constraining $\Delta$ to be less than $\pi$
comes from the fact that $B = A + \pi k/n < A + \pi$ since $k < n$.
This in particular excludes indiscernible points located at the zeros of the cosine
function as well as indiscernible points of periodicity of $2\pi$ or higher.
Combining the previous expressions, we then have $\Delta= \pi k/n$
Thus discernibility is violated whenever exist $k,i,n$ and $m$ such that 
$A = \pi k i /n - \pi k/2n =   \pm m \pi - \pi k/2n$
or, equivalently, 
\begin{eqnarray} \label{eq:ex}
k i = n m
\end{eqnarray}
where $n >1$. The following conclusions can be drawn:
i) A trivial solution to (\ref{eq:ex}) is given by $k=m=0$.
As previously noted, this 
corresponds to the fact that the consensus state $x = \mathbf{1}$
turns out to be indiscernible for any link/node disconnection;
ii) Nontrivial solutions to (\ref{eq:ex}) also exist. Simple examples 
are $(k,i,n,m)=(2,4,8,1)$ and $(k,i,n,m)=(8,5,10,4)$; 
iii) Apart from the case where $k=m=0$, 
there is no solution to (\ref{eq:ex}) when $i \in \{1,n-1\}$. In fact, 
if $i=$ we obtain $k  = n m$. This has no solutions 
when $k,m>0$ since $k < n$. If instead $i=n-1$ we obtain 
$k (n-1) = n m$, or, equivalently, $k- m = k/n$.  This has no solutions 
when $k,m>0$ since $k-m$ is integer, whereas $k/n$ 
cannot be integer since $k < n$. 

Point iii) is interesting since it indicates that, apart from the stationary state 
$x = \mathbf{1}$, 
a link disconnection can always be detected if it involves  one of the endpoints 
of the graph. The latter situation can also be viewed as a node disconnection,
and, in fact, it is straightforward to verify that, apart from the stationary case,
node disconnections are always detectable.
To see this, notice that, 
in accordance with Theorem 4, a
node disconnection is not detectable if and only if 
there exist an index $i$ such that
{\setlength\arraycolsep{0pt} 
\begin{eqnarray*} 
cos(\pi k i /n - \pi k/2n) \, &=& \,  cos(\pi k i /n + \pi k/2n) \\
&=& \,  cos(\pi k i /n - 3\pi k/2n)
\end{eqnarray*}}%
By the point iii) above, one can restrict the attention
to the case where $i \in \{3,4,\ldots,n-2\}$ since the 
cases $i \in \{1,2\}$ and $i \in \{n-1,n\}$ do involve
the endpoints of the graph. Let $A$ and $B$ be as before, and let $C := \pi k i /n - 3 \pi k/2n$.
Thus, there must exist three points, namely 
$A,B$ and $C$ where the cosine function takes on the 
same value. 
Apart from the case $k=0$, this is not possible since 
$B=C+2\pi k/n<C+2\pi$. Hence, we conclude that, 
apart from the stationary case, a node disconnection 
is always detectable.

It is worth mentioning that a very similar analysis could be carried out 
with respect to the grid graph since its eigenspace 
is completely determined by the eigenspace of the path graph \cite{Notarstefano}.

\section{Conclusions}

In this paper, we have addressed the problem of detecting topology variations in dynamical networks, considering 
both the cases of node and link disconnections. 
The results show that the detection problem can
be characterized through simple algebraic conditions, which depend on the eigenspace components related to 
the nominal and faulty operating mode of the network. 
While the analysis is mainly oriented towards a 
theoretical characterization of the detection problem,
the results also provide several insights on how detection 
can be addressed in practice, as well as guidelines 
for the development of sensor placement algorithms.

\section*{Appendix}

{\em Proof of Theorems 1-4.} Let $x$ be an eigenvector of $\Phi$ with eigenvalue $\lambda$, 
\emph{i.e.}, $x$ is such that $\Phi x = \lambda x$.
Then, Theorem 1 readily follows from the fact that, when a link disconnection without dynamics
reconfiguration occurs, we have
\[
\bar \Phi x = \Phi x - \phi_{ij} (e_i \, e'_j + e_j \, e_i') x  = \lambda x - \phi_{ij} (x_j \, e_i + x_i \, e_j).
\]
Similarly, in the case of a link disconnection with dynamics
reconfiguration we have
\begin{eqnarray*}
\bar \Phi x &=& \Phi x + \phi_{ij} (e_i \, e'_i + e_j \, e_j' + e_i \, e'_j + e_j \, e_i') \, x \\
&=& \lambda x + (x_i - x_j) \, e_i + (x_j -x_i) \, e_j
\end{eqnarray*}
from which Theorem 2 follows. As for the case of a node disconnection without dynamics reconfiguration,
we have
{\setlength\arraycolsep{0pt} 
\begin{eqnarray*}
\bar \Phi x \, &=& \, \Phi x - \sum_{j \in \mathcal N_i} \phi_{ij} \, (e_i \, e'_j + e_j \, e_i') \, x \\
&=& \, \lambda x - \left (  \sum_{j \in \mathcal N_i} \phi_{ij}  \, x_j \right ) e_i - x_i \,  \sum_{j \in \mathcal N_i} \phi_{ij} \, e_j
\end{eqnarray*}}%
which leads to the statement of Theorem 3. Finally, if we consider the case of a node disconnection with dynamics reconfiguration, we have
{\setlength\arraycolsep{0pt} 
\begin{eqnarray*}
\bar \Phi x \, &=&\, \Phi x + \sum_{j \in \mathcal N_i} (e_i \, e'_i + e_j \, e_j' + e_i \, e'_j + e_j \, e_i') \, x \\
&=& \, \lambda x - \left (  \sum_{j \in \mathcal N_i} (x_i- x_j) \right ) e_i + \sum_{j \in \mathcal N_i} (x_j - x_i) \, e_j
\end{eqnarray*}}%
and, hence, condition (\ref{eq:prop4}). 
Concerning the second part of Theorem 4, notice that when 
$\Phi = -L$ then the $i$-th
row of the condition $\Phi  x = \lambda x$ can be written as
\[
-  | \mathcal N_i | \, x_i +  \sum_{j \in \mathcal N_i}  x_j = \lambda \, x_i .
\]
Thus, it can be seen that $x_j = x_i $ for any $j \in \mathcal N_i$ implies $\lambda \, x_i = 0$ which concludes the proof.
\qedp

{\em Proof of Theorem 5.} 
Clearly, conditions (i) and (ii) are necessary because, otherwise, we would have non-null initial states leading to zero
output trajectories and hence indiscernible from the output.
Further, condition (i) ensures that for any state $x$ such that $S' (\lambda) x \ne 0$ with  $\lambda \in
{\rm spec} (\Phi) \setminus {\rm spec} (\bar \Phi)$ the term $e^{\lambda t} M S (\lambda) S' (\lambda)$ related to the output evolution
of $(\mathcal G, \Phi)$ is not null. Since this term is never present in the output evolution of $(\bar {\mathcal G}, \bar \Phi)$,
we can conclude that $(\mathcal G, \Phi)$ is $M$-discernible from $(\bar {\mathcal G}, \bar \Phi)$ in such a state.
Similarly, condition (ii) ensures that $(\bar {\mathcal G}, \bar \Phi)$  is $M$-discernible from $(\mathcal G, \Phi)$ in all the
states $x$ such that $\bar S' (\lambda) x \ne 0$ for at least one  $\lambda \in
{\rm spec} (\bar \Phi) \setminus {\rm spec} ( \Phi)$. Hence, under conditions (i) and (ii), a necessary condition
for a pair of states $(x, \bar x)$ to be $M$-indiscernible is that  $S' (\lambda) x = 0$ and $\bar S' (\bar \lambda) \bar x = 0$
for any $\lambda \in
{\rm spec} (\Phi) \setminus {\rm spec} (\bar \Phi)$ and $\bar \lambda \in {\rm spec} (\bar \Phi) \setminus {\rm spec} (\Phi)$,
or, equivalently, that $x$ belongs to $\cup_{\lambda \in {\rm spec} (\Phi) \cap {\rm spec} (\bar \Phi)} V(\lambda)$ 
and  $\bar x$ belongs to $\cup_{\lambda \in {\rm spec} (\Phi) \cap {\rm spec} (\bar \Phi)} \bar V(\lambda)$.
Consider now two such states $x$ and $\bar x$. Condition (iii) ensures that, for any  
$\lambda \in {\rm spec} (\Phi) \cap {\rm spec} (\bar \Phi)$, one has $M S(\lambda) S'(\lambda) x = 
M \bar S(\lambda) \bar S'(\lambda) \bar x$ if and only if $S(\lambda) S'(\lambda) x = \bar S(\lambda) \bar S'(\lambda) \bar x$.
This, in turn, implies that $(x,\bar x)$ are $M$-indiscernible if and only if $x = \bar x \in \mathcal I$, which proves the
sufficiency of conditions (i)-(iii). Finally, the necessity of condition (iii) readily follows from the fact that, 
when such a condition is not satisfied for some $\lambda \in {\rm spec} (\Phi) \cap {\rm spec} (\bar \Phi)$, 
there exist pairs of states 
$(x, \bar x) \notin \mathcal I_{\rm P}$ 
with $x \in V (\lambda)$, $\bar x \in \bar V (\lambda)$
such that $M S(\lambda) S'(\lambda) x = 
M \bar S(\lambda) \bar S'(\lambda) \bar x$. \qedp

{\em Proof of Lemma 1.} 
Consider the joint observability matrix $[\mathcal O_N \; -\bar{\mathcal O}_N]$ which coincides with the standard
sampled-data observability matrix associated with the pair $(\Gamma, \Delta)$. Since ${\rm spec} (\Delta) = 
{\rm spec} (\Phi) \cup {\rm spec} (\bar \Phi)$ and $\Delta$ has dimension $2 n \times 2 n$, we can apply the 
Kalman-Bertram criterion for the observability of sampled-data systems and conclude that, 
when conditions (\ref{eq:obs:sample}) and $N \ge 2n$ are satisfied, the null space of 
$[\mathcal O_N \; -\bar{\mathcal O}_N]$
coincides with the set of unobservable states of the pair $(\Gamma, \Delta)$, which is precisely the set $\mathcal I (M)$. 
Hence, we have $\mathcal O_N x = \bar{ \mathcal O}_N \bar x$ if and only if $(x, \bar x) \in \mathcal I (M)$. 
Then, the statement of Lemma 1 can be proven by proceeding as in the proof of Theorem 2 of \cite{battSCL}, to which the reader is referred for details.
\qedp

\bibliographystyle{elsarticle-num}

\bibliography{Connection_topology}

\end{document}